\RequirePackage{fix-cm}
\documentclass[smallcondensed]{svjour3}      % onecolumn (ditto)
\smartqed  % flush right qed marks, e.g. at end of proof
\usepackage{graphicx}
\usepackage{subfigure}
\usepackage{amssymb}
\usepackage{amsmath}
\usepackage{mathrsfs}
\usepackage{placeins}
\usepackage{natbib}
\usepackage{caption3}
\bibpunct{[}{]}{,}{a}{,}{,}

\usepackage[bookmarksnumbered=true,breaklinks=true,colorlinks,citecolor=blue,linkcolor=blue]{hyperref}
\makeatletter

\newcommand{\numtoRoman}[1]{\expandafter\@slowromancap\romannumeral #1@}

\renewcommand{\figurename}{Figure}
\usepackage[usenames]{color}

\makeatother

\begin{document}

\title{On the underlying drag-reduction mechanisms of flow-control strategies in a transitional channel flow: temporal approach}

\author{Alexander J. Rogge         \and
        Jae Sung Park}

\institute{J. S. Park \at
              Department of Mechanical and Materials Engineering, University of Nebraska-Lincoln, Lincoln, NE 68588-0526, USA 
              \email{jaesung.park@unl.edu}
}

\date{Received: date / Accepted: date}
% The correct dates will be entered by the editor

\maketitle

\begin{abstract}
The underlying mechanisms of three different flow-control strategies on drag reduction in a channel flow are investigated by direct numerical simulations at friction Reynolds numbers ranging from 65 to 85. These strategies include the addition of long-chain polymers, the incorporation of slip surfaces, and the application of an external body force. While it has been believed that such methods lead to a skin-friction reduction by controlling near-wall flow structures, the underlying mechanisms at play are still not as clear. In this study, a temporal analysis is employed to elucidate underlying drag-reduction mechanisms among these methods. The analysis is based on the lifetime of  intermittent phases represented by the active and hibernating phases of a minimal turbulent channel flow (Xi \& Graham, $\textit{Phy. Rev. Lett.}$ 2010). At a similar amount of drag reduction, the polymer and slip methods show a similar mechanism, while the body force method is different. The polymers and slip surfaces cause hibernating phases to happen more frequently, while the duration of active phases is decreased. However, the body forces cause hibernating phases to happen less frequently but prolong its duration to achieve a comparable amount of drag reduction. A possible mechanism behind the body force method is associated with its unique roller-like vortical structures formed near the wall. These structures appear to prevent interactions between inner and outer regions by which hibernating phases are prolonged. It should motivate adaptive flow-control strategies to exploit the distinct underlying mechanisms for robust control of turbulent drag at low Reynolds numbers.

%Bursting phenomenon can be understood in terms of the instability of the nonlinear traveling wave solutions to the Navier-Stokes equations.

% \PACS{PACS code1 \and PACS code2 \and more}
% \subclass{MSC code1 \and MSC code2 \and more}
\end{abstract}

\vspace*{+0.3in}

\section{introduction}

The presence of coherent structures in wall-bounded turbulent flows plays a crucial role in turbulent dynamics \cite{RobinsonARFM1991}. Near the wall, these structures are closely related to the self-sustaining process of turbulence as they are very similar to staggered, counter-rotating quasi-streamwise vortices, forming low- and high-speed streaks \cite{Waleffe1997}. In particular, these near-wall structures are responsible for the production of turbulent kinetic energy as they are observed to burst in a very intermittent fashion \cite{Hamilton1995}. Since the bursting process appears to account for over 80\% of the energy in turbulent fluctuations, these near-wall coherent structures are believed to be the dominant structures associated with the turbulence production, leading to an increase in skin-friction drag \cite{Lumley1998}. Thus, various control strategies have been exploited to manipulate the near-wall coherent structures to achieve a significant drag reduction \cite{Gad2007}. Although the gross effects and structural understanding of drag-reduction mechanisms of the flow-control strategies have been well-documented, a temporal analysis on elucidating the underlying drag-reduction mechanisms of the control strategies has yet to be explored until now.

Here, we aim to provide a brief description of the control strategies that have been explored along with their implications in drag-reduction mechanisms derived by such strategies. In general, the control strategies are classified as passive or active, depending on whether it requires actuation or external energy source \cite{Gad2007}. For passive control, one of the most successful strategies involves using riblets or microgrooves installed on the wall and aligned in the streamwise direction \cite{Choi1993,Jimenez2004}. Drag reduction varies depending on the height and alignment of the riblets, resulting in up to 10\% drag reduction. The riblets enforce the streamwise vortices away from the wall by which the amount of shear stress in the near-wall region is significantly reduced \cite{Choi1993, Garcia2011}. Another substantially-studied strategy is via slip by placing hydrophobic surfaces at the walls \cite{Luchini1991,Watanabe1999,Min2004b}. Recently, superhydrophobic surfaces, which are a combination of surface roughness and surface chemistry at the micro- and/or nano-scales, have been utilized to produce an effective slip length on the wall \cite{RothsteinARFM2010,Fukagata2006}. It is viewed that the effective slip length must be on the same order as the viscous sublayer to alter the streamwise velocity and subsequently the wall shear stress fluctuations \cite{Park2013d}. A reduction in the wall shear stress results in the weakening of the streamwise vortices and streaks, which subdues the streamwise momentum to move away from the wall, which is also known as the lift-up mechanism. Since the lift-up mechanism is reduced by the weakening of the wall shear stress, the transient growth of perturbations is also reduced \cite{Chai2019}. Another most successful strategy involves the addition of a small amount of long-chain polymers to a liquid \cite{Berman1978}. The polymers are likely to store elastic energy and release it back into the flow as it travels around the buffer and log-law layers, whereby the streamwise vortices are suppressed \cite{White2008,Graham2014}. The vortices tend to remain closer to the wall but become elongated in the direction of the flow. A substantial drag reduction up to 80\% can be achieved, yielding a much higher flow rate at a given pressure drop \cite{Graham2014}. For active control, one of the most practical strategies involves active wall motion, which creates streamwise ridges whose appearance is very similar to that of riblets \cite{Choi2002, Quadrio2004, Kang2000}. This wall motion also attempts to push the high-speed fluid away from the wall to obtain drag reduction \cite{Endo2000}. Another popular strategy is the application of blowing and suction at the wall, which is equal or opposite to the wall-normal velocity close to the wall \cite{Choi1994}. This control strategy, also known as opposition control, resulted in a drag reduction of nearly 25\%. The opposition control reduces the spinning of the streamwise vortices and in turn stabilizes them in space, leading to a reduction of the bursting frequency \cite{Coller1994}.  Another method of active control includes the application of external body forces to the flow, which can be applied to both gases and liquids \cite{Karniadakis2003,Berger2000,Wang2013}. The spanwise external body forces appear to stabilize the low-speed streaks and weaken streamwise vortices, leading to the weakening of the bursting events. As a summary, almost all of the various control strategies appear to suggest that their underlying drag-reduction mechanisms are to manipulate near-wall coherent structures, such as streamwise vortices or streaks near the wall. 

Besides the structural drag-reduction mechanisms, there have been very limited research to compare different flow-control strategies to elucidate different or similar drag-reduction mechanisms. Very recently, Chen, Yao, \& Hussain have utilized the energy-box analysis \cite{Gatti2018} to compare the three control methods of the spanwise opposed wall-jet forcing (SOJF), spanwise wall oscillation (SWO), and opposed wall blowing/suction (OBS) at a friction Reynolds number of 200 and drag reduction of approximately 20\% \cite{Chen2021}. They compared the contributions of the mean, coherent, and random turbulent dissipations to the overall drag reduction and net power savings. It was found that for the SOJF method, the coherent dissipation is much smaller than the other two dissipations, while for the SWO method, the coherent value is on the same scale as the other two values and much larger than SOJF`s value. For the OBS method, the random turbulent dissipation is suppressed without the appearance of the coherent dissipation since the energy is only introduced randomly. In addition to the energy-box analysis, it should be noted that there are other approaches to connect the drag to certain flow quantities, such as the Fukagata-Iwamoto-Kasagi (FIK) identity \cite{Fukagata2002} and the Renard-Deck (RD) identity \cite{Renard2016}. A few examples of exploiting the FIK identity for different flow control methods are wall deformation \cite{Tomiyama2013}, external body forces \cite{Mamori2014}, superhydrophobic surfaces \cite{Lee2015}, and blowing/suction \cite{Kametani2015}, among others. Although these approaches provide quantitatively meaningful information about drag-reduction mechanisms of different flow-control methods, there is still demand to better understand underlying drag-reduction mechanisms of different flow-control strategies via distinct approaches.

In addition to turbulent flow control for drag reduction, a wall-bounded turbulent flow itself exhibits substantial intermittency between high- and low-drag states. Particularly relevant in this regard is the study by Xi \& Graham \cite{Xi2010}, where a direct numerical simulation (DNS) was performed for a minimal turbulent channel flow at a friction Reynolds number of 85. Turbulent dynamics are observed to cycle intermittently between ``active" intervals with strong streamwise vortices and ``hibernating" intervals with very small Reynolds shear stress. Similar observations have also been made by Hamilton, Kim, \& Waleffe \cite{Hamilton1995}. This temporal intermittency is also found to reflect the organization of the turbulent dynamics around the exact coherent states or nonlinear traveling-wave (TW) solutions to the Navier-Stokes equations \cite{Park2015jfm,Park2018,Graham2021}. Indeed, the hibernating intervals are approaches to lower-branch TW solutions, while the active intervals are close to an upper-branch TW solution. Very recently, this low- and high-drag intermittency is observed and quantified in experiments and in good agreement with DNS results \cite{Rishav2020}. It is worth noting that the temporal and spatial analyses on the relationship between temporal dynamics in minimal domains and spatiotemporal dynamics in extended domains also yield very similar results for the hibernating and active intervals \cite{Kushwaha2017} at friction Reynolds numbers ranging from 70 to 100. While the connection between the temporal intermittency and drag-reduction mechanisms has been identified for viscoelastic turbulent flows at low Reynolds numbers \cite{Graham2014}, it has yet to be fully explored for other control strategies and will be investigated in the present study. {Along with the polymer method, the current study examines the slip surface method (passive control) and external body force method (active control).} The description of such a connection would provide a basis for a deeper understanding of underlying drag-reduction mechanisms embedded in different flow-control strategies.

In this paper, we use direct numerical simulations and temporal turbulent phases to elucidate the underlying mechanisms of drag reduction via three control strategies, namely the application of polymer additives, slip surfaces, and external body forces. The problem formulation is reported in Section \ref{sec:formulation}. The simulation results are presented in Section \ref{sec:results}, where the effects of the control strategies on the temporal intermittency at the different levels of drag reduction are presented. A summary of main results, conclusions, and future directions are presented in Section \ref{sec:conclusion}.

\section{Problem Formulation}\label{sec:formulation}
We consider an incompressible fluid in a turbulent channel flow (plane Poiseuille) geometry, driven by a constant volumetric flux $Q$. The $x$, $y$, and $z$ coordinates are aligned with the streamwise, wall-normal, and spanwise directions, respectively. Periodic boundary conditions are imposed in the $x$ and $z$ directions with fundamental periods $L_x$ and $L_z$, and solid walls are placed at $y=\pm h$, where $h$ is the half-channel height. The laminar centerline velocity for a given volumetric flux is given as $U_c = (3/4)Q/h$. Using the half-height $h$ of the channel and the laminar centerline velocity $U_c$ as the characteristic length and velocity scales, respectively, the nondimensionalized Navier-Stokes equations for a fluid velocity $\boldsymbol{u}$ and pressure $p$ are then given as
\begin{equation} \label{eq:ns}
\nabla \cdot \boldsymbol{u}=0,\quad
\frac{\partial \boldsymbol{u}}{\partial t}+\boldsymbol{u}\cdot\nabla \boldsymbol{u}=-\nabla p+\frac{\beta}{Re_c}\nabla^2\boldsymbol{u} + \boldsymbol{f}_{\textrm{ext}}. %\frac{1-\beta}{Re_c Wi}(\nabla\cdot \boldsymbol{\tau}_{p}) + \boldsymbol{f}_b.
\end{equation}
Here, we define the Reynolds number for the given laminar centerline velocity as $Re_c = U_c h/\nu$, where $\nu$ is the kinematic viscosity of the fluid and $\beta$ is the ratio of the solvent viscosity and the total viscosity (for a Newtonian fluid, $\beta = 1$), and $\boldsymbol{f}_{\textrm{ext}}$ is the external force, which can result from a body force or polymer stress in the present study.

For viscoelastic flows, the momentum equation in Equation (\ref{eq:ns}) includes an external force from the polymer stress $\boldsymbol{f}_{\textrm{ext}} = (1-\beta) \nabla \cdot \boldsymbol{\tau}_p$, where the polymer stress tensor $\boldsymbol{\tau}_p$ is related to the polymer conformation tensor $\boldsymbol{\alpha}$. This tensor is then expressed through the FENE-P constitutive relation based on bead-spring dumbbells. These polymer conformation and stress tensors are obtained by solving the following equations:
\begin{align}
\frac{\partial \boldsymbol{\alpha}}{\partial t}+ \boldsymbol{u}\cdot \nabla \boldsymbol{\alpha}-\boldsymbol{\alpha}\cdot \nabla \boldsymbol{u} -(\boldsymbol{\alpha}\cdot \nabla \boldsymbol{u})^{\textrm{T}} = -\frac{1}{Wi}\boldsymbol{\tau}_p, \quad \boldsymbol{\tau}_p = \frac{\boldsymbol{\alpha}}{1-\textrm{tr}(\boldsymbol{\alpha})/b}-\textbf{\textsf{I}}.
\end{align}
Here, we define the Weissenberg number $Wi = \lambda U_c/h$, where $\lambda$ is the polymer relaxation time, and $b$ is the maximum extensibility of the polymers. For the current study, we fix $\beta = 0.97$ and $b = 10,000$. Since $1-\beta$ is proportional to polymer concentration and $b$ to the number of monomer units, this parameter set corresponds to a dilute solution of a high-molecular-weight polymer. For slip surfaces, streamwise Navier slip conditions $u_{s} = L_{s} \dot{\gamma}_{w}$ are applied at both top and bottom walls, where $L_s$ is an effective homogeneous slip length and $\dot{\gamma}_{w}$ is the shear rate at the wall. For an external body force, the following spanwise body force is used for the external force term in equation (\ref{eq:ns}):
\begin{align}
f_{z}=I \textrm{e}^{-y/\Delta} \textrm{sin} \left(\frac{2\pi}{\lambda_{z}}z-\frac{2\pi}{T}t\right),
\end{align}
where $I$ is the amplitude of excitation, $\Delta$ is the penetration depth, and $\lambda_{z}$ and $T$ are the wavelength and period of oscillation, respectively. For the present simulations, we vary $I$ and $T$, while fixing $\Delta = 0.03$ and $\lambda_z = L_z/2 = \pi/2$.

For inner units, characteristic inner scales are the friction velocity $u_{\tau}=(\bar{\tau}_w/\rho)^{1/2}$ and the near-wall length scale or wall unit $\delta_{\nu} = \nu/u_{\tau}$, where $\rho$ is the fluid density and $\bar{\tau}_w$ is the time- and area-averaged wall shear stress. As usual, quantities nondimensionalized by these inner scales are denoted with a superscript ``+". The friction Reynolds number is then defined as $Re_{\tau}=u_{\tau}h/\nu=h/\delta_{\nu}$.

Simulations are performed using the open-source code \textit{ChannelFlow} written and maintained by Gibson \cite{channelflow} from which a modified version was made for the three different control methods used in the current study. {In this study, we focus on the domains of $L_x \times L_y \times L_z = 2\pi \times 2 \times \pi$, utilizing the minimal flow unit (MFU) approach \cite{Jimenez1991}.} {Computational verification of direct numerical simulations for MFUs has been tested up to $Re_{\tau} = 1000$ \cite{Davis2021}. The minimum spanwise length scale used here is about 205$\delta_{\nu}$, which is larger than the length scale of the near-wall streak spacing of about $100\delta_{\nu}$ \cite{Smith1983}.} {In particular. a grid convergence and domain dependence have been tested in our previous studies for no-control cases \cite{Kushwaha2017,Rishav2020} and in the current study for control cases.} A numerical grid system is generated on $N_x \times N_y \times N_z$ (in $x$, $y$, and $z$) meshes, where a Fourier-Chebyshev-Fourier spectral spatial discretization is applied to all variables. A typical resolution used is $(N_x, N_y, N_z) = (48, 81, 48)$. The numerical grid spacing in the streamwise and spanwise directions are $\Delta x_{min}^{+} \approx 11.0$, $\Delta z_{min}^{+} \approx 5.5$. The nonuniform Chebyshev spacing used in the wall-normal direction results in $\Delta y_{min}^{+} \approx 0.05$ at the wall and $\Delta y_{max}^{+} \approx 2.5$ at the channel center. For simulations, $Re_c = 1800$ is being considered, which gives $Re_{\tau}$ = 85 for no control and slightly lower $Re_{\tau}$ values for control methods due to a lower value of $u_{\tau}$. Prior to presenting simulation results, it should be noted that all simulations presented are sufficiently far above transition and show the sustained turbulence nature of the flow, yielding the classical mean flow behavior \cite{Kushwaha2017,Whalley2017}, which will be revisited below.

\section{Results and discussion}\label{sec:results}
\subsection{Drag reduction by control strategies}\label{sec:dr}

\begin{figure}
 \begin{center}\includegraphics[width=1\textwidth]{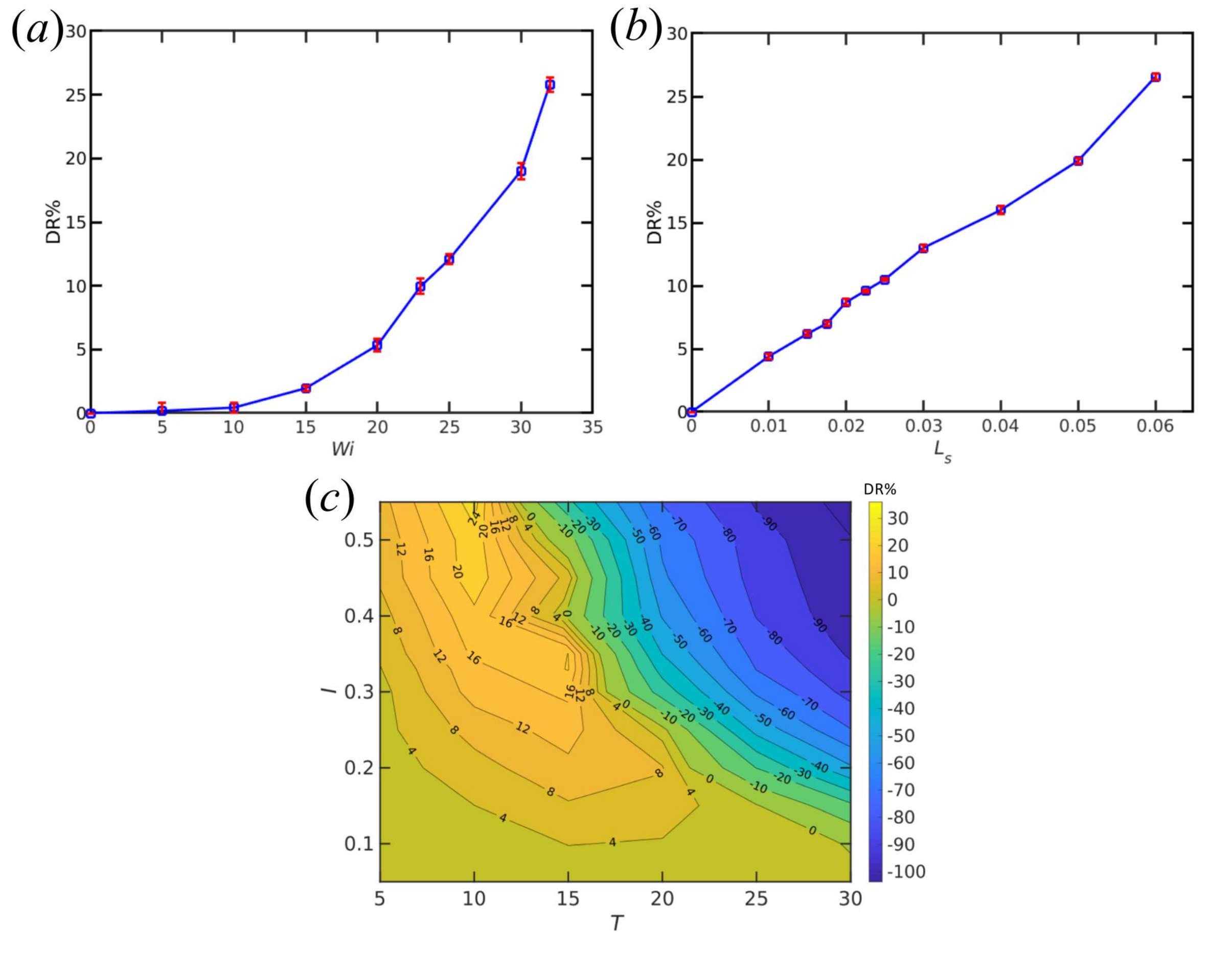}
  \caption{Drag reduction percentages (DR\%) due to ($a$) polymer additives, ($b$) slip surfaces, and ($c$) spanwise body forces. The error bars in ($a$) and ($b$) are computed by the block-averaging method to represent the standard error of the time-averaged quantity \cite{Flyvbjerg1989}.
    \label{fig:dr}}
    \end{center}
\end{figure}

Figure~\ref{fig:dr} shows the level of drag reduction due to polymer additives (i.e. viscoelastic flows), slip surfaces, and spanwise body forces. Drag reduction percentage is defined as DR\% = $(C_{f,un} - C_f)/C_{f,un} \times 100\%$, where $C_f = 2 \overline{\tau}_w /(\rho U^2_b)$ is the skin-friction coefficient for the controlled case and $C_{f,un}$ is the friction factor for the uncontrolled case. Here, $\overline{\tau}_w$ is the time-area-averaged wall shear stress and $U_b$ is the bulk fluid velocity, which is kept constant in simulations. The error bars on the plot are the standard errors of the time-averaged quantity with the block-averaging method \cite{Flyvbjerg1989}. Figure~\ref{fig:dr}($a$) shows DR\% for viscoelastic turbulence as a function of $Wi$, which is in good agreement with the previous studies with regard to drag reduction amounts and onset $Wi$ for drag reduction \cite{Xi2010,Wang2014}. Figure~\ref{fig:dr}($b$) shows DR\% for slip surfaces as a function of $L_s$. At a fixed Reynolds number, the slip length and drag reduction percentage are almost linearly correlated, which is also observed with the same streamwise-only slip condition of Min and Kim \cite{Min2004b}. It is worth noting that the largest {slip length value} is $L_s^+ \approx 5$ for the current study, which ensures that the homogeneous slip surface employed in the present study would produce essentially the same outcomes with a heterogeneous microtextured slip surface or superhydrophobic surface \cite{Seo2018,Picella2019,Davis2020,Rowin2019}. Figure~\ref{fig:dr}($c$) shows DR\% for body forces for various values of the amplitude of excitation ($I$) and the time-period of oscillation ($T$), where we fix $\Delta = 0.03~(\Delta^+ = 2.55$) and $\lambda_z = L_z/2 = \pi/2~(\lambda_z^+ = 42.5)$. Within parameters studied, the maximum drag reduction percentage is approximately 25\% at $I = 0.55$ and $T = 10~(T^+ = 40)$ -- the values in parentheses are based on no-control case.

\begin{figure}
 \begin{center}\includegraphics[width=1\textwidth]{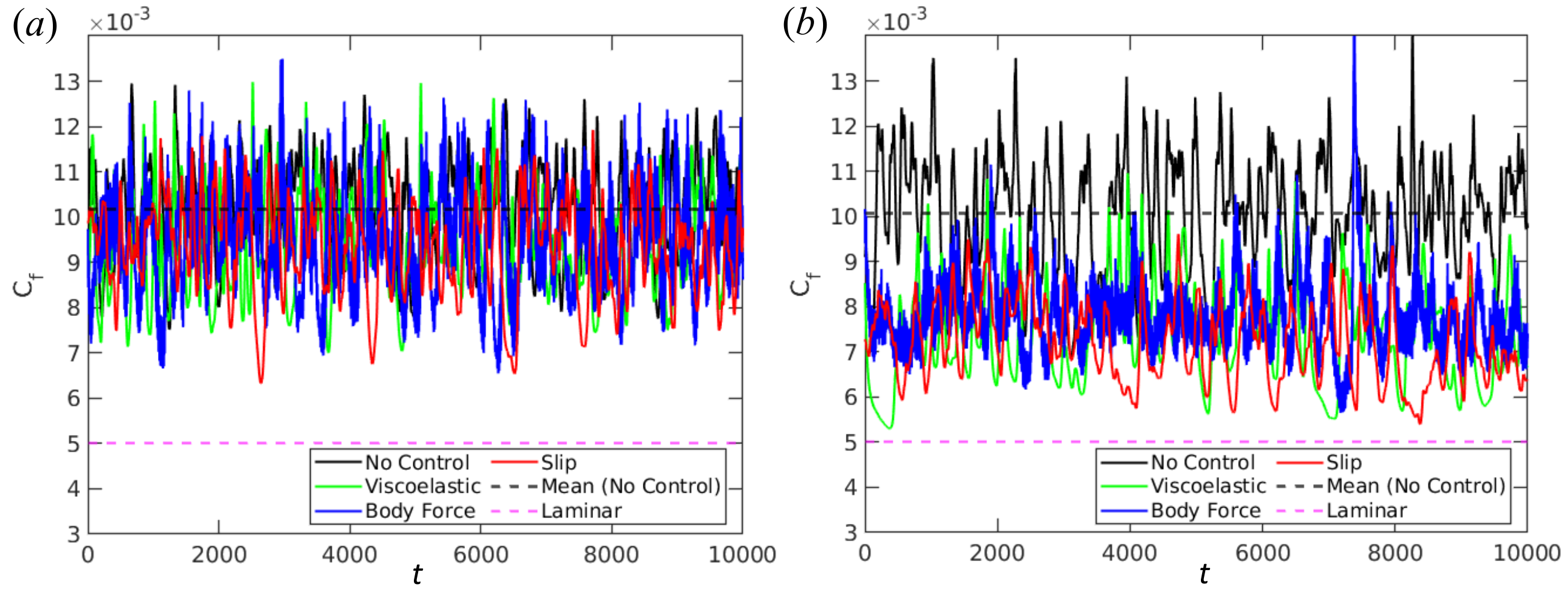}
  \caption{Time series of skin friction coefficients $(C_f)$ at about ($a$) 7\% and ($b$) 25\% drag reduction for viscoelastic (green), slip (red), and body force (blue) methods along with no control (black), its mean value (black dashed), and laminar (pink dashed).
    \label{fig:cf}}
    \end{center}
\end{figure}

Before moving forward, it should be again emphasized that the simulations indeed show characteristics of a sustained turbulence. Figures \ref{fig:cf}($a$) and ($b$) show a time series of the skin friction coefficients for no control and three control methods along with laminar value at DR\% $\approx$ 7\% and 25\%, respectively. {The parameters used for the control methods at DR\% $\approx$ 7\% include $Wi = 20$ for the viscoelastic method, $L_s$ = 0.015 for the slip method, and $I = 0.15$ and $T = 15$ for body force method. At DR\% $\approx$ 25\%, $Wi = 31$ for the viscoelastic method, $L_s$ = 0.06 for the slip method, and $I = 0.55$ and $T = 10$ for the body force method.} As shown, the turbulent nature of the flow is clearly visible, showing the substantial fluctuations, which are zero in a laminar flow by definition. The skin friction coefficients are well above the laminar for low drag reduction cases (Fig. \ref{fig:cf}$a$). Although it tends to approach the laminar value as drag reduction increases, it is still above the laminar state and shows noticeable fluctuations even during the lowest drag periods of DR\% $\approx$ 27\% at which the maximum DR case is achieved {in the current study} with a slip surface. Even in this period, there are no quasi-laminar regions in the simulation domain that every region always displays the fluctuations.
%The velocity fluctuations are also significant during these periods, remaining at roughly the same value when normalized in wall units (not shown) \cite{Whalley2017}.

\vspace*{-0.1in}

\subsection{Low- and high-degree drag reduction regimes}\label{sec:ldrhdr}

\begin{figure}
 \begin{center}\includegraphics[width=1\textwidth]{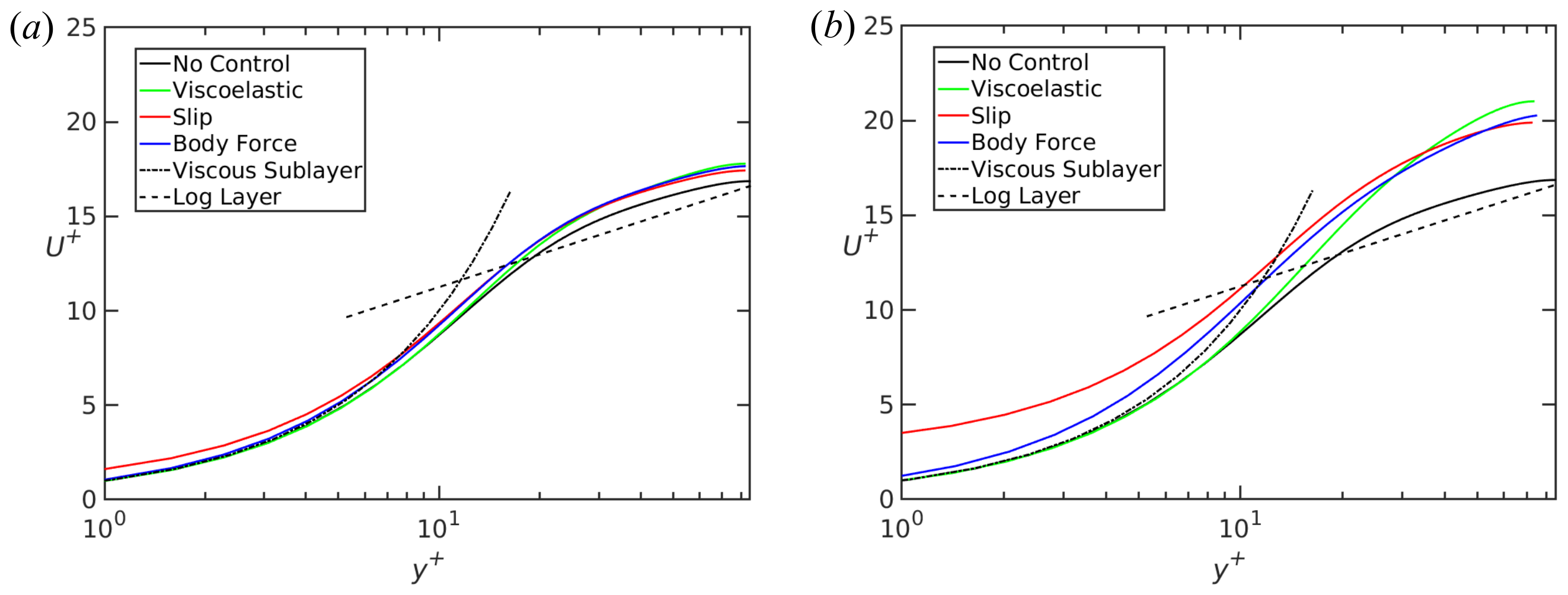}
  \caption{Time-averaged streamwise mean-velocity profiles for no control (black), viscoelastic (green), slip (red), and body force (blue) methods at about ($a$) 7\% and ($b$) 25\% drag reduction. The black dot-dashed line is the viscous sublayer, $U^+ = y^+$ and the black dashed line is the log-law layer, $U^+ = 2.5 \textrm{ln}(y^+) + 5.5$.
    \label{fig:meanvel}}
    \end{center}
\end{figure}

Prior to proceeding to different drag-reduction regimes, it is worth showing the mean velocity profiles, which can be used as an indicator for different levels of drag reduction. Figures~\ref{fig:meanvel}($a$) and ($b$) show the mean velocity profiles (uncontrolled and controlled) in inner units at DR\% $\approx$ 7\% and 25\%, respectively. {The control parameters are the same as ones for Figure~\ref{fig:cf}}. For comparison, the profiles for the viscous sublayer $U^+(y^+) = y^+$ and the log-law layer $U^+ = 2.5 \textrm{ln}(y^+) + 5.5$ are also presented. For DR\% $\approx$ 7\%, the polymer and body force methods follow the viscous sublayer profile well $(y^+ \le 5)$, while the slip case starts off with a greater velocity than the other two methods due to the slip velocity at the wall. The control profiles begin to slightly deviate from the no-control profile in the buffer layer at $y^+ \approx 20$. The no-control profile lies closer to the log-law profile, but it is placed just above the log-law profile because of the effects of low Reynolds number \cite{Tsukahara2005}. The velocity profiles of three control methods are elevated beyond the no-control and are very close to one another, suggesting almost the same lower drag. For DR\% $\approx$ 25\%, the mean velocity profiles of the polymer and body force methods clearly diverge from the log-law slope with a steeper incline. The slip profile shows similar values with the other two cases in the log-law layer but with a less steep incline. However, there is a much-increased velocity at the wall due to higher $L_s$.

Regarding the drag reduction percentages, an additional distinction can be applied to separate a low degree of drag reduction (LDR) and a high degree of drag reduction (HDR). Warholic $\textit{et al.}$ analyzed the different effects of polymers at LDR and HDR regimes through experiments based on the mean velocity and fluctuation profiles relative to Newtonian turbulent flows \cite{Warholic1999}. This difference has been recently investigated through simulations to show various ranges for LDR and HDR regimes, where LDR is below 13\% drag reduction and HDR is above 15\% drag reduction \cite{Zhu2019}. A similar separating point for the LDR-HDR transition was  observed in the range of DR\% $\approx$ 20\%--30\% \cite{Xi2010}. Although some studies reported a higher separating point at DR\% $\approx$ 30\%--40\% \cite{Patasinski2003,Min2003}, the present drag reduction data can be divided into the LDR and HDR regimes at DR\% $\approx$ 15\%--20\% based on the mean profiles, as seen in fig.~\ref{fig:meanvel}. Thus, the 7\% and 25\% drag reduction cases belong indeed to the LDR and HDR regimes, respectively, where the steeper mean velocity profile slope distinguishes the HDR regime from the LDR regime. %\Note{We did see 10\% and 20\% as well, and the results are almost identical -- we should mention this somewhere} Only before or after the transitional region will the results be similar, however the results begin to differentiate significantly when compared across this transition region.
\vspace*{-0.1in}
\subsection{Temporal analysis on drag-reduction mechanisms}\label{sec:mech}
We now aim to describe the underlying drag-reduction mechanisms based on a temporal characterization of temporal events with varying amounts of drag relative to the mean. It has been seen that a turbulent flow is observed to intermittently fluctuate between low and high friction drag. During a low-drag period, vortical motions are suppressed with less wavy low-speed streaks, causing low Reynolds shear stress \cite{Graham2014}. These low-drag intervals are termed hibernating turbulence \cite{Xi2010}. Periods between the hibernating intervals are called active turbulence and display high-drag features. The criteria for hibernating turbulence involve the wall shear stress remaining below $90\%$ of its mean value for at least three eddy turnover times $( > \Delta t u_{\tau}/h = 3)$ \cite{Kushwaha2017,Rishav2020}. We already addressed the issue of sensitivity to the chosen values and showed almost identical results \cite{Kushwaha2017,Park2018}. Note that with the aforementioned criteria for hibernating turbulence, it can be only detected up to $Re_{\tau} \approx 125$ \cite{Whalley2017}.

To quantify the effects of the flow-control methods on the temporal intermittency of turbulence, the duration and frequency of hibernating and active turbulence are computed from simulation runs for $t = 150000$ $(> 80 Re_c)$ for all cases. The average duration of hibernating and active turbulence and the fraction of time spent in hibernation ($T_H$, $T_A$, $F_H$, respectively) are calculated as
\begin{align}
T_{H} = \frac{\sum_{i=1}^{N_{H}} t_{H,{i}}}{N_{H}},\; \; \; T_{A}= \frac{T - \sum_{i=1}^{N_{H}} t_{H,{i}}}{N_{A}},\; \; \; F_{H}= \frac{\sum_{i=1}^{N_{H}} t_{H,{i}}}{T},
\end{align}
where $t_{H,i}$ is the duration of the $i$th hibernating interval, and $N_{H}$ and $N_{A}$ are the total number of hibernating and active intervals over the total duration of the simulation $T$, respectively. Specifically, $F_H$ can be referred to as a temporal intermittency factor for low friction drag \cite{Kushwaha2017}. For temporal investigations, these three quantities $T_H$, $T_A$, and $F_H$ are considered along with the average number of hibernations $n_H$ over $t = 30000$ for the LDR and HDR regimes in comparison to no-control case.

\begin{figure}
 \begin{center}\includegraphics[width=1\textwidth]{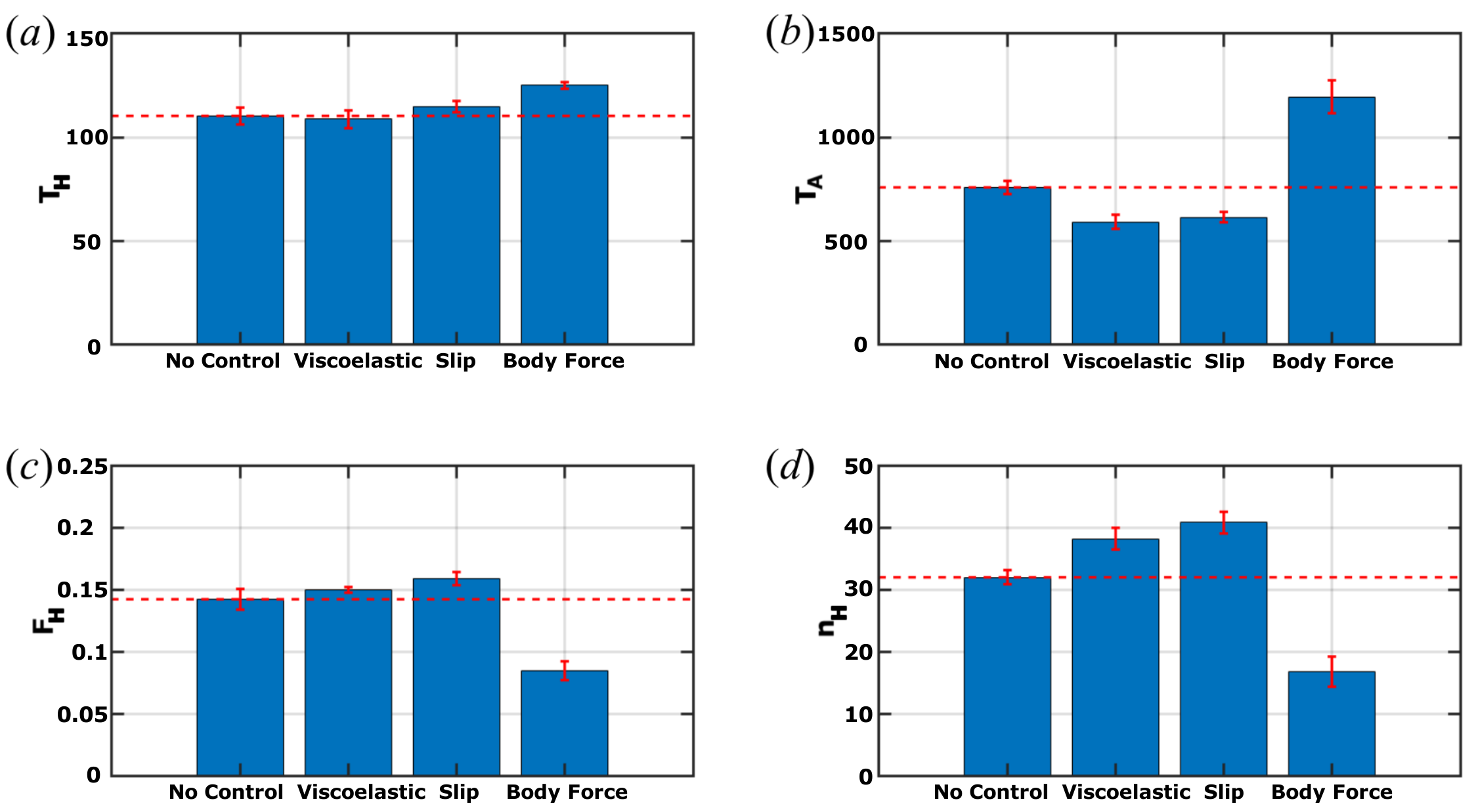}
  \caption{Temporal data at DR\% $\approx$ 7\% (LDR): ($a,b$) the average duration of hibernating and active turbulence, respectively, ($c$) the fraction of hibernation, and ($d$) the average number of hibernations over $t = 30000$. The results are from viscoelastic ($Wi = 20$), slip ($L_s$ = 0.015), and body force ($I = 0.15, T = 15$) cases. The dashed (red) line represents the no-control case's values for better visualization of changes that occur. The error bars represent the standard error.
    \label{fig:bar7}}
    \end{center}
\end{figure}

Figure \ref{fig:bar7} shows the temporal quantities for a drag reduction of $7\%$ at the LDR regime. It appears that the polymer and slip methods share similar characteristics. They show almost the same duration of hibernating intervals as the no-control case but a decrease in the duration of active intervals. These changes lead to an increase in the fraction of hibernation and the number of hibernations compared to the no-control case. These trends are in good agreement with the previous studies on viscoelastic turbulence at low $Wi$ \cite{Xi2010,Wang2014}. However, the body force method shows different behaviours, where the duration of both hibernating and active intervals increases, while the fraction of hibernation and the average number of hibernations decrease compared to the no-control values. These trends might indicate that the body force method could cause a drag increase rather than a drag reduction of $7\%$ even though $T_H$ is larger than the no-control value. Thus, it is strongly suggested that the body force method is likely to have a different drag-reduction mechanism compared to the polymer and slip methods at the LDR regime. {Before moving to HDR, it should be noted that given the almost same drag reduction percentage of 7\%, the temporal quantities for the body force method with $(I = 0.15, T = 15)$, $(I = 0.15, T = 20)$, and $(I = 0.2, T = 10)$ are almost identical (not shown), indicating the independence of the temporal analysis the chosen control values for the body force method.}

\begin{figure}
 \begin{center}\includegraphics[width=1\textwidth]{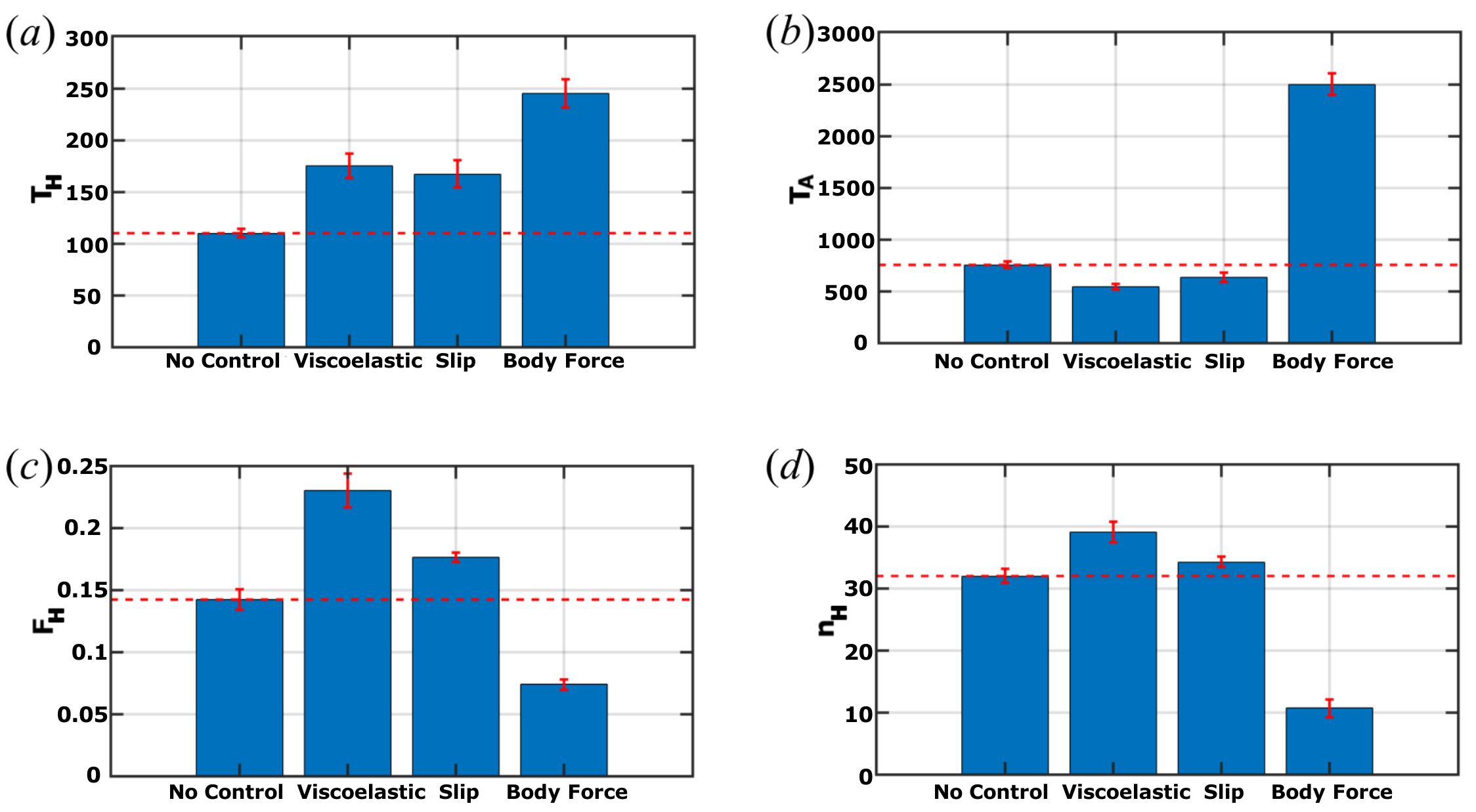}
  \caption{Temporal data at DR\% $\approx$ 25\% (HDR): ($a,b$) the average duration of hibernating and active turbulence, respectively, ($c$) the fraction of hibernation, and ($d$) the average number of hibernations over $t = 30000$. The results are from viscoelastic ($Wi = 31$), slip ($L_s$ = 0.06), and body force ($I = 0.55, T = 10$) cases . The dashed (red) line represents the no-control case's values for better visualization of changes that occur. The error bars represent the standard error.
    \label{fig:bar25}}
    \end{center}
\end{figure}

Figure \ref{fig:bar25} shows the cases for a drag reduction of $25\%$ at the HDR regime. As in the LDR regime, the polymer and slip control methods display similar behaviours. While the trends of $T_A$, $F_H$, and $n_H$ with respect to the no-control case are similar to the LDR cases, the average duration of hibernating turbulence does increase and is now larger than the no-control value. With this increase in $T_H$ and the resulting increase in $F_H$, a much higher drag reduction ($\sim$25\%) is achieved. For the body force method, the trend is still similar to its LDR case but shows more noticeable changes in the quantities. In comparison to the LDR case, $T_H$ and $T_A$ become almost doubled, while $F_H$ remains almost the same, and $n_H$ decreases slightly. These trends still might indicate that the body force method could cause a drag increase rather than a drag reduction of $25\%$. Thus, a different drag-reduction mechanism could be suggested for the body force method compared to the polymer and slip methods even at the HDR regime.

With the temporal quantities at the LDR and HDR regimes, the underlying drag-reduction mechanisms can be made for the control methods being investigated. As seen above, it appears that the mechanism is almost the same for polymer and slip methods. For LDR, they achieve a drag reduction by decreasing the duration of active intervals, while the duration of hibernating turbulence remains almost constant. These trends lead to more frequent hibernation and an increase in the temporal intermittency factor. For HDR, while they still show a decrease in the duration of active turbulence, more drag reduction is achieved by increasing the duration of hibernating turbulence, which results in an increase in $F_H$ compared to the LDR case. Interestingly, the body force method displays a distinctly different mechanism at both LDR and HDR regimes. Although it causes less frequent hibernation and even a smaller $F_H$ value than the no-control case, the highly prolonged hibernation intervals are likely to produce comparable amounts of drag reduction to the polymer and slip methods.

\begin{figure}
 \begin{center}\includegraphics[width=1\textwidth]{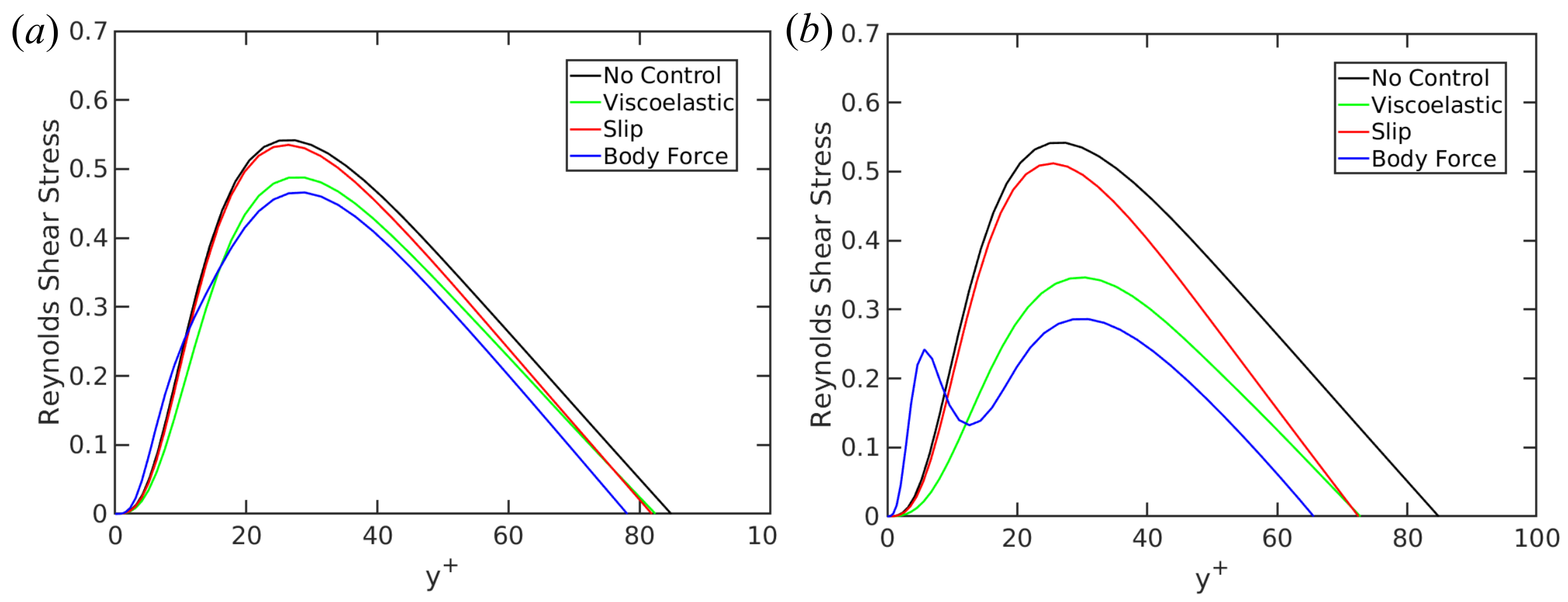}
  \caption{($a,b$) Reynolds shear stress profiles in inner units for the flow-control methods at about 7\% and 25\% drag reduction, respectively, along with no-control case.
    \label{fig:rss}}
    \end{center}
\end{figure}

\vspace*{-0.1in}
\subsection{Mechanisms behind different flow-control strategies}\label{sec:vortex}
Lastly, we attempt to illuminate the mechanisms behind the different temporal characteristics for drag reduction between the polymers/slip methods and the body force method. {As alluded by different polymer dynamics during active and hibernating turbulence in viscoelastic flows \cite{Graham2014}, the polymer and slip methods appear to possess a selective turbulent preventative mechanism. For LDR, as seen in Figure 4, the polymers and slip surfaces seem to affect only the duration of the active turbulence phases, while not making any changes to the duration of the hibernating turbulence phases. It suggests that the effects of polymers and slip surfaces are amplified with turbulence-induced shear. In other words, the drag-reducing mechanism actively kicks in when turbulence is excited, such as during active turbulence. The polymers begin to stretch and store elastic energy to release it back into the flow. The slip velocity or slip length becomes larger in inner units when wall shear stress gets higher. Again, this selective mechanism is noticeable during LDR as the low-drag states are unchanged and the high drag states are only affected. For HDR, as seen in Figure 5, it is still observed that the selective mechanism affects the high-drag states as done for LDR. However, the low-drag states are now also affected as the hibernating turbulence appears to take over turbulent dynamics as more drag reduction is achieved \cite{Graham2014,Wang2014}. A detailed investigation of drag-reduction mechanisms during HDR will be included in future work.}

It has been shown that the effects of flow-control methods on the Reynolds shear stress and vortical structures may provide a mechanistic basis for drag reduction in turbulent flows \cite{Lumley1998,Gad2007}. Figures~\ref{fig:rss}($a$) and ($b$) depict the time-averaged Reynolds shear stress profiles of the different flow-control methods along with a no-control case for the LDR and HDR regimes, respectively. For LDR, the three control profiles are slightly lower than the no-control case. Interestingly, it is observed that the body force profile is slightly higher than the polymers/slip and no-control profiles in $y^+ < 15$. For HDR, the three control profiles are fairly reduced compared with the magnitudes of the no-control profile. The slip profile is still relatively close to the no-control case compared to the other two cases. More interestingly, a change in the shape of the body force profile is noticeable, which is non-monotonic. Mostly, the profile is rather higher than the no-control case in $y^+ < 10$, with the peak being close to $y^+ \approx 7$, and its magnitude falls below the other cases from $y^+ > 18$, with another peak at almost the same locations as the other profiles. Note that this non-monotonic shape is kept even for hibernation and active intervals (not shown), while the magnitudes are different.

\begin{figure}
 \begin{center}\includegraphics[width=1\textwidth]{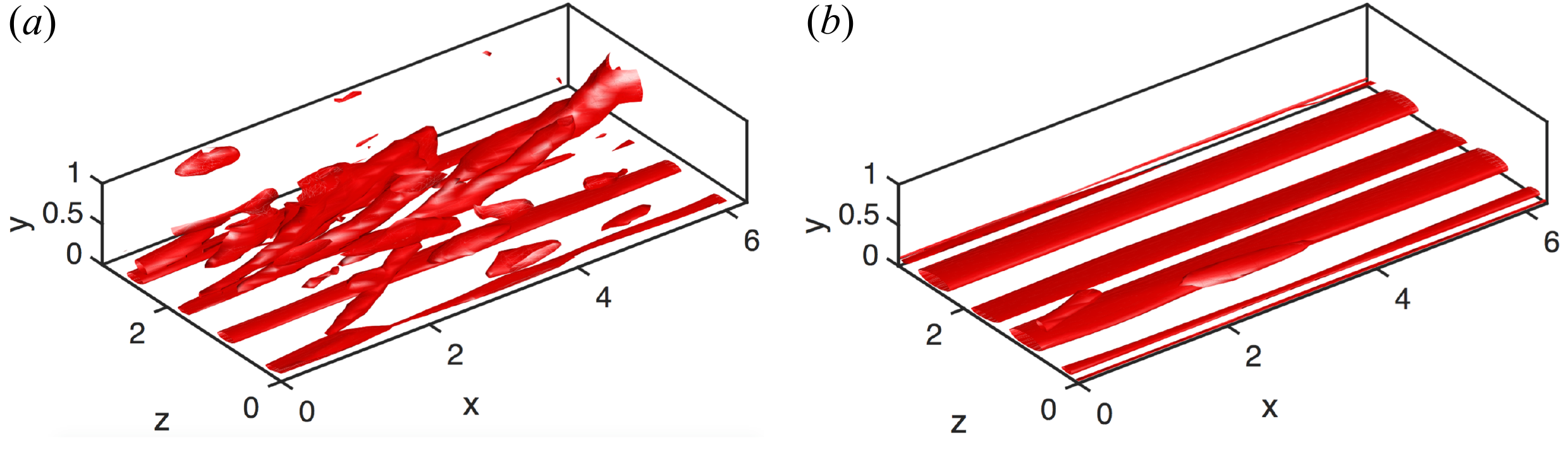}
  \caption{($a,b$) Vortical structures due to the body force method for LDR (DR = 7\%) and HDR (DR = 25\%) regimes, respectively. The red tubes are isosurfaces of 1/2 of maximum swirling strength $\lambda_{ci}$, which corresponds to the imaginary part of the complex conjugate eigenvalues of the velocity gradient tensor \cite{ZhouJFM1999}.
    \label{fig:vortex}}
    \end{center}
\end{figure}

To elucidate the seemingly distinct characteristics of the body force method for a drag-reduction mechanism, we plot its vortical structures for which the swirling strength $\lambda_{ci}$, the imaginary part of the complex conjugate eigenvalues of the velocity gradient tensor \cite{ZhouJFM1999}, are calculated. {Prior to presenting the vortical structures due to the body force, it should be noted that the main effects of the polymers and slip surfaces on flow structures are well-documented that vortices are weakened \cite{xi2010jfm,Davis2020}.} Figures~\ref{fig:vortex}($a$) and ($b$) show contours of the swirling strength for half of the channel for the LDR and HDR regimes, respectively, {during the hibernating phases due to the body force method.} The contours represent isosurfaces of 50\% of the maximum $\lambda_{ci}$ for each case. Similar to ones found by Mamori \& Fukagata \cite{Mamori2014}, the roller-like structures are clearly observed very close to the wall and formed in the streamwise direction. {However, it should be mentioned that Mamori \& Fukagata had the roller-like structures formed in the spanwise direction due to the Lorentz force applied in the wall normal direction.} As the wavelength of the body force is set to $L_z/2$, four roller-like vortices are formed and separated by $L_z/4$ -- a pair of vortices per wavelength. It is observed that the heights of these vortices are independent of the choice of the other control parameters such as penetration depth ($\Delta$), force magnitude ($I$), and period ($T$). The center of the rollers is located at $y^+ \approx 7$. Given these streamwise-spanned roller-like vortices, it can be speculated that the body force is likely to prevent interactions between the inner region and outer region by which the hibernation intervals tend to be substantially prolonged. In other words, these streamwise roller-like structures tend to stay near the wall and block structures from the outer region, which helps increase the hibernation intervals for the reduction of friction drag. The observation in vortical structures due to body forces could tie into a shear sheltering effect, which effectively limits an interplay between the inner and outer layers \cite{hunt1999}. {As drag reduction is further increased, the effect of the body force becomes more noticeable as it causes to form a strong shear sheltering layer in a buffer layer, which effectively decouples the flow structures above and below this layer. This blocking mechanism is different from the selective mechanism of the polymer and slip methods.}

The differences in Reynolds shear stress and vortical structures between the polymer/slip methods and the body force method may provide clear and plausible mechanisms responsible for distinct underlying drag-reduction mechanisms, for which further investigation is yet needed. In particular, further investigation is need to understand how vortical structures separate the flow regions due to a body force. In addition, as there have been limited studies to relate a shear sheltering to drag reduction mechanisms \cite{ptasinski2003}, a detailed connection of the body force method to a shear sheltering will be a subject of interesting future work.

\section{Conclusion}\label{sec:conclusion}
Through direct numerical simulations in a channel flow (plane Poiseuille) geometry at low (transitional) Reynolds numbers in a range of $Re_{\tau} = 65 - 85$, the underlying drag-reduction mechanisms of three flow-control strategies, namely polymer additives, slip surfaces, and external body forces, are investigated by utilizing two temporal turbulent phases -- hibernating and active turbulence \cite{Xi2010}. Given similar drag-reduction percentages, the polymers/slip methods demonstrate a similar drag-reduction mechanism to one another by causing the hibernation phases to occur more frequently with a decrease in the duration of active phases. The body force method shows a different mechanism, where the hibernation phases happen less frequently, however the duration of these hibernation phases is prolonged due to roller-like vortical structures formed near the wall. These vortical structures appear to prevent interactions between the inner and outer regions to make hibernating phases longer. At higher drag-reduction regimes, each control method seems to involve different mechanisms to manipulate the Reynolds shear stress for which follow-up work is under investigation. {More importantly,  a connection of the temporal analysis to the drag-reduction mechanisms due to the three flow control methods has been made and suggests that the polymer and slip methods possess a ‘selective’ turbulent preventative mechanism, affecting only active turbulence, and the body force method possesses a ‘blocking’ mechanism, affecting both active and hibernating turbulence.}

%\color{purple}The paper by Graham \cite{Graham2014} mentions that the viscoelastic control method could posses a selective mechanism; during the active turbulence of LDR, polymers are only affecting the average duration of the active intervals while not making any changes to the hibernating interval duration. In the present study it appears that the slip control method also contains a selective mechanism that only affects the duration of active intervals during periods of active turbulence. The reason the body force case is different from the other methods is that its mechanism is not selective and instead affects both hibernating and active turbulence. The selective mechanisms show up during LDR where the low drag states are unchanged and only affect the high drag areas. During HDR, it is not as clear since the mechanism begins affecting low drag states as well as the periods of high drag. \color{black}

These distinct underlying drag-reduction mechanisms between the polymers/slip methods and body force method could suggest that different adaptive and optimal flow-control techniques could be used to promote more drag reduction, which will be a subject of interesting future work. For instance, when a flow enters a hibernation phase with an aid of an external body force, a special control strategy could be applied at the locations of the streamwise roller-like structures to stabilize them to make a hibernation phase much longer for more drag reduction. In addition, a further study is necessary to investigate the effect of higher Reynolds numbers on the current temporal approach for a practical relevance in practice. However, it should be noted that hibernation phases are barely detected beyond $Re_{\tau} \approx 125$ \cite{Whalley2017} using the current hibernation criteria. Thus, the criteria should be somewhat relaxed for higher Reynolds number flows to detect hibernating turbulence. This research avernue is currently under investigation \cite{Davis2020dfd,Ryu2021}. Finally, the temporal approach of the current study can be combined with the quantatiative approaches such as the Fukagata-Iwamoto-Kasagi (FIK) identity \cite{Fukagata2002}, Renard-Deck (RD) identity \cite{Renard2016}, and energy-box analysis \cite{Gatti2018} to provide more comprehensive information about the underlying drag-reduction of different flow-control strategies.

\begin{acknowledgements}
The direct numerical simulation code used was developed and distributed by J. Gibson at the University of New Hampshire. The authors also acknowledge the computing facilities used at the Holland Computing Center at the University of Nebraska-Lincoln.
\end{acknowledgements}

\noindent
\textbf{Funding} The authors gratefully acknowledge the financial support from the National Science Foundation through a grant OIA-1832976, the Nebraska EPSCoR FIRST Award supported by the National Science Foundation through a grant OIA-1557417, and the Collaboration Initiative at the University of Nebraska.

\section*{Declarations}
\textbf{Conflict of interest}  The authors declare that they have no conflict of interest.
\\
\\
\noindent
\textbf{Availability of data and material} The data that support the findings of this study are available from the corresponding author upon reasonable request.

% Authors must disclose all relationships or interests that 
% could have direct or potential influence or impart bias on 
% the work: 
%\section*{Conflict of Interest}
%The authors declare that they have no conflict of interest.

% BibTeX users please use one of
%\bibliographystyle{spbasic}      % basic style, author-year citations
%\bibliographystyle{spmpsci}      % mathematics and physical sciences
%\bibliographystyle{spphys}       % APS-like style for physics
\bibliographystyle{ieeetr}

\bibliography{refs}
\end{document}